\documentclass[usenatbib]{mn2e}

\usepackage[pdftex]{graphicx}
\usepackage{aas_macros}
\usepackage{amsfonts,amssymb}
\usepackage[pdftex,
    a4paper,
    plainpages=false,
    pdfpagelabels,
    breaklinks=true,
    bookmarks=true,
    bookmarksopen=true,
    bookmarksopenlevel=1]{hyperref}

\title{Phase-Space Density Profiles in Scale-Free Cosmologies}
\author[S.R.~Knollmann, A.~Knebe and Y.~Hoffman]
       {Steffen~R.~Knollmann$^1$, Alexander~Knebe$^1$ and Yehuda~Hoffman$^2$\\
        $^1$Astrophysikalisches Institut Potsdam, An der Sternwarte 16,
            14482 Potsdam, Germany\\
        $^2$Racah Institute of Physics, Hebrew University, Jerusalem
            91904, Israel}

\begin{document}
\maketitle
\begin{abstract}
  We use a set of high-resolution simulations of scale-free Einstein-de Sitter
  cosmologies to investigate the logarithmic slope of the phase-space
  density profile $Q(r) = \rho(r)/\sigma^3(r)$ of dark matter (DM) haloes.
  The initial conditions for the simulations are determined by a power
  law power spectrum  of the form $P(k) \propto k^n$.  We compute the
  $Q(r)$ profiles using the radial, tangential and full velocity dispersion,
  and the velocity anisotropy parameter, $\beta(r)$. We express $Q(r)$ as
  a single power-law $Q(r) \propto r^\alpha$ and derive a median slope
  $\alpha$ in each simulation and for each definition of $Q$.  Our main
  findings are: 1. The various $Q(r)$ profiles follow a power law to a
  good approximation.  2. The slopes depend on the concentration parameter
  $c$ of the DM haloes, where for $c \gtrsim 10$ the slopes steepen with
  rising concentration and for $c \lesssim 10$ the trend flattens and
  even turns around.  3. The asymptotic value of $\beta$ as $r\rightarrow
  R_{\mathrm{vir}}$ increases with the value of $c$.  4. In accordance
  with \citet{Zait2007} $\alpha_{\mathrm{rad}}$ becomes more negative as
  the asymptotic value of $\beta$ at the virial radius increases.  5.
  This introduces a weak dependence of the $Q(r)$ slopes on the
  slope of the power spectrum.
\end{abstract}



\section{Introduction}

For the density profile $\rho(r)$ of dark matter (DM) haloes a variety of
empirical fitting formulae exist.  The one applied mostly is the
so-called NFW \citep{Navarro1996,Navarro1997} profile, which gives a
fair description for haloes found in a wide range of $N$-body
simulations.  Recent studies though are hinting to use a different
functional form for the density profile that differs from the usual
NFW form mainly in the central regions \citep[see,
e.g.][]{Navarro2004,Merritt2006}, but common to all suggested profiles
is the fact that they are not described by a simple power-law but
rather require a smoothly changing logarithmic slope.  It comes
therefore as a surprise that the coarse-grained phase-space
density\footnote{See, e.g., \citet{Dehnen2005} for clarifications
about the applicability of the term 'phase-space density', as the
term is commonly used, however, we will continue to refer to $Q(r)$
as the phase-space density profile.} profile, calculated from the
density- and the velocity dispersion profile $\sigma(r)$ as
\begin{equation}
\label{eq:qr}
Q(r) = \frac{\rho(r)}{\sigma^3(r)},
\end{equation}
very closely follows a {\em single} power-law $Q(r) \propto r^\alpha$
with $\alpha \approx -1.9$.  This was first noted by
\citet{Taylor2001} and has since been confirmed by many other works
\citep[e.g.][]{Boylan-Kolchin2004, Rasia2004, Ascasibar2004, Austin2005,
Dehnen2005, Hoffman2007, Ascasibar2008}.

Different velocity dispersions have been used with
equation~\ref{eq:qr}, generally the radial velocity dispersion
$\sigma_{\mathrm{rad}}(r)$ is used, but also the total velocity
dispersion $\sigma_{\mathrm{tot}}(r)$ has been employed.  Both lead to
separate definitions for $Q$ that nevertheless can be well fitted with
a single power-law, albeit marginal differences in the slopes
\citep{Dehnen2005, Faltenbacher2007, Ascasibar2008}, see also the
discussion in~\citet{Zait2007}.

The power law phase-space density profile has been found mostly in
simulations of structure formation in the standard model of cosmology,
namely the flat $\Lambda$CDM model.  It is not known how general its
behaviour in other cosmologies is and the dynamical origin of the
power-law is not yet understood.  However, the structure of (spherical)
haloes in equilibrium obeys the (spherical) Jeans equation which relates
the density profile, the phase-space density profile and the velocity
anisotropy parameter $\beta$ (to be defined below).  As mentioned above,
the universality of the density profile is well-established across a
large variety of cosmological models.  The description of the
phase-space density profiles as a single power-law, however, is not that
well checked.

Hence we aim here at studying the (phase-space) structure of DM haloes
in different cosmogonies, in particular scale-free Einstein-de Sitter
cosmologies with varying spectral index $n$, to explicitly probe
universality of the phase-space profile.  We use the numerical
simulations of \mbox{\cite{Knollmann2008}} who studied the structure of haloes
in pure DM flat cosmologies with a primordial power spectrum of the form
$P(k) \propto k^n$, finding the equilibrium haloes to be well described
by a universal density profile.  Scale-free models provide a clean
probe of the possible dependence of the structure of DM halos on the
primordial initial conditions.  This provides insight  to the
concordance $\Lambda$CDM model in the sense that the power spectrum for
this cosmology exhibits a varying slope depending on the scale.

We shall further investigate the
behaviour of the phase-space density profile with different components
of the total velocity dispersion $\sigma_\mathrm{tot}$, namely the
radial and tangential dispersions.  Given the close connection between
the $Q(r)$ power law and the velocity anisotropy \citep{Zait2007},
expressed by the $\beta$ parameter, the $\beta$
profile of the DM haloes is to be explicitly studied. 
The paper is organized as follows. The numerical experiments are briefly
discussed in \S 2. The fitting of the phase-space density profile is
described  in \S 3 and the results are presented in \S 4. The paper
concludes with a summary and discussion in \S 5.

\section{Numerical Experiments}

We base our analysis on the simulations described
in~\citet{Knollmann2008}.  Those simulations are set up according to a
power spectrum of the initial density perturbations of the form
\begin{equation}
  P(k) \propto k^{n}.
\end{equation}
Five different models have been simulated with
$n = -0.50$, $-1.50$, $-2.25$, $-2.50$ and $-2.75$.  The simulations were performed
with the parallel $N$-body code \textsc{Gadget2}~\citep{Springel2005}
following the evolution of $512^3$ dark matter particles.  The final
stage of each run is chosen in such a way that $M_*$ -- the typical
collapsing mass -- reaches the same value for all choices of
$n$~\citep[for a more detailed description see][]{Knollmann2008}.

\begin{table}
  \caption{Simulation details.}
  \label{tab:simus}
  \begin{tabular}{@{}clcccccc}
    \hline
    Name     & Type & $B$ & $A$/$\sigma_8$ & $\Omega_m$ & $\Omega_\Lambda$ & $N$  & $N_h$ \\
    \hline
    512-0.50 & $n=-0.50$    &  $1$ & $0.0358$ & $1.0$ & $0.0$ & $512^3$ & $186$ \\
    512-1.50 & $n=-1.50$    &  $1$ & $0.0215$ & $1.0$ & $0.0$ & $512^3$ & $106$ \\
    512-2.25 & $n=-2.25$    &  $1$ & $0.0109$ & $1.0$ & $0.0$ & $512^3$ & $61$  \\
    512-2.50 & $n=-2.50$    &  $1$ & $0.0076$ & $1.0$ & $0.0$ & $512^3$ & $34$  \\
    512-2.75 & $n=-2.75$    &  $1$ & $0.0046$ & $1.0$ & $0.0$ & $512^3$ & $14$  \\
    256-B50  & $\Lambda$CDM & $50$ & $0.9$    & $0.3$ & $0.7$ & $256^3$ & $19$  \\
    \hline
  \end{tabular}

	\medskip
    Summary of the simulations. The first column simply
    assigns a unique label used throughout the paper while the second
    column specifies the model. $B$ refers to the side length of the
    simulation box. $A$ ($\sigma_8$) is the normalisation of the input
    power spectrum used to generate the initial conditions. $\Omega_m$
    and $\Omega_\Lambda$ describe the background cosmology. $N$ gives
    the number of particles in the model and $N_h$ the number of
    haloes used for the analysis.
\end{table}

We then employ the \textsc{MPI} version of the \textsc{AMIGA} Halo
Finder\footnote{\textsc{AMIGA} is freely available for download at
  \url{http://www.aip.de/People/AKnebe/AMIGA/}} (\textsc{AHF},
successor of \textsc{MHF} introduced by \citet{Gill2004}) to identify
haloes and to compute their radial and integral properties.  The
radial properties are calculated in logarithmic bins in $r$ and only
those bins which are converged according to the criteria presented in
\citet{Power2003} are retained for the analysis.  We further restrict
our sample of haloes to those having a mass $M$ in the range $[0.75
M_*, 1.5 M_*]$ where a typical $M_*$ halo is resolved with $4.2 \times
10^4$ particles.

Also, we require the haloes to be relaxed, for which we have used the
displacement $\Delta_r$ of the centre of mass of all material inside the
virial radius with respect to the potential centre normalized to the
virial radius 
\begin{equation}
  \Delta_r = \left|\mathbf{r}_{\mathrm{cm}} -
              \mathbf{r}_{\mathrm{cen}} \right|/r_\mathrm{vir}.
\end{equation}
For inclusion in the sample, we require $\Delta_r \le 0.05$, employing a
more conservative value than proposed by other recent
studies~\citep[e.g.][]{Neto2007}.

For comparison we also use the results of a ``standard'' $\Lambda$CDM
simulation analysed at $z=0$ whose particulars are described in detail
elsewhere (Power et al. in prep.). Using that simulation we
define a halo sample labelled 256-B50 including all haloes resolved
with more than $3.15 \times 10^4$ particles (corresponding to a halo
of mass $0.75 M_*$ in the scale-free simulations) and being relaxed
according to the same ``off-centre'' criterion alluded to above.  A
summary of the simulation parameters and the halo samples is given in
table~\ref{tab:simus}.

\section{Phase-Space Density Profiles}

The evaluation of the phase-space density as defined by
equation~\ref{eq:qr} requires the (total) velocity
dispersion $\sigma_\mathrm{tot}(r)$ at the radius $r$, which is
calculated as
\begin{equation}
\label{eq:veldisp}
  \sigma_{\mathrm{tot}}^2 (r)
 = \frac{1}{N_\mathrm{sh} - 1}
   \sum_{i=1}^{N_\mathrm{sh}}
   \left| \mathbf{v}_{\mathrm{tot},i} -
          \left\langle\mathbf{v}_\mathrm{tot}\right\rangle
   \right|^2
\end{equation}
where the sum extends over the $N_\mathrm{sh}$ particles contained in
a shell with radius $r$. With $\mathbf{v}_{\mathrm{tot},i}$ and
$\left\langle\mathbf{v_\mathrm{tot}}\right\rangle$ we refer to the
velocity of the $i$th particle and the average velocity within the
shell, respectively.  For the calculation of the radial phase-space
density $Q_{\mathrm{rad}}(r)$ we project the 3D velocity
$\mathbf{v}_{\mathrm{tot},i}$ onto the radial direction and use only
the radial component with equation~\ref{eq:veldisp}, respectively.
With these two velocity dispersions we can obtain the tangential
velocity dispersion $\sigma_{\mathrm{tan}}$ by using the relation
\begin{equation}
\label{eq:sigmarel}
  \sigma_{\mathrm{tan}}^2 = \sigma_{\mathrm{tot}}^2 -
\sigma_{\mathrm{rad}}^2
\end{equation}
for the calculation of the tangential phase-space density profile
$Q_{\mathrm{tan}}(r)$.

\subsection{Normalization}

For each halo we normalize the radius $r$ by $r_{2}$ and the $Q$
profile by $Q_{2} \equiv Q(r_{2})$, where $r_{2}$ is the radius at
which the density profile has a slope of $-2$.\footnote{In the case of
an NFW profile $r_2$ corresponds to the scale radius $r_s$.} We find
this ``scaling radius'' in a non-parametric way by locating the
maximum of $r^2 \rho(r)$.  Having identified $r_2$ for each halo and
representing its $Q$-profile with a cubic spline, $Q_{2}$ is then
readily available as the evaluation of the spline interpolation at
$r_2$.

Additionally, we define for each halo a concentration parameter $c$
as
\begin{equation}
  c = \frac{r_\mathrm{vir}}{r_{2}}
\end{equation}
where $r_\mathrm{vir}$ is the virial radius defined via
$3M(<r_{\mathrm{vir}})=4\pi\Delta_{\mathrm{vir}}\rho_b
r^3_{\mathrm{vir}}$ with $\rho_b$ being the cosmic background density
and $\Delta_{\mathrm{vir}}$=178 (340) for the scale-free
($\Lambda$CDM) model. The concentration $c$ has been found to be a
distinguishing quantity between haloes from different scale-free
models, with haloes forming in a shallower cosmology being more
concentrated~\citep[cf.][]{Knollmann2008}.

\subsection{Fitting}

The normalized $Q$-profiles of each halo are fitted by a single power-law
\begin{equation}
q(r; \alpha) = Q_2 \left(\frac{r}{r_2}\right)^{\alpha}
\end{equation}
by minimizing the quantity
\begin{equation}
\label{eq:chisqr}
\chi^2(\alpha) = \sum_i\left(\frac{Q(r_i) - q(r_i; \alpha)}{0.1 \cdot
Q(r_i)}\right)^2.
\end{equation}
Note that due to the choice of our normalization we have only the
slope $\alpha$ as a free parameter.

\section{Results}

In this section, we will describe our results, starting in
subsection~\ref{subsec:qprofs} with the power-law nature of the
phase-space density profiles.   We will then, in
subsection~\ref{subsec:universality}, focus on the individual slopes
probing for the generality of the slopes and on what quantities they
depend. We will then shortly investigate the relation between the three
different $Q$ profiles in subsection~\ref{subsec:relation_q}, before
finally focussing on the velocity anisotropy profiles in
subsection~\ref{subsec:betaprofs}.

\subsection{Phase-space density profiles}
\label{subsec:qprofs}

\begin{figure*}
\begin{center}
  \includegraphics[width=\textwidth]{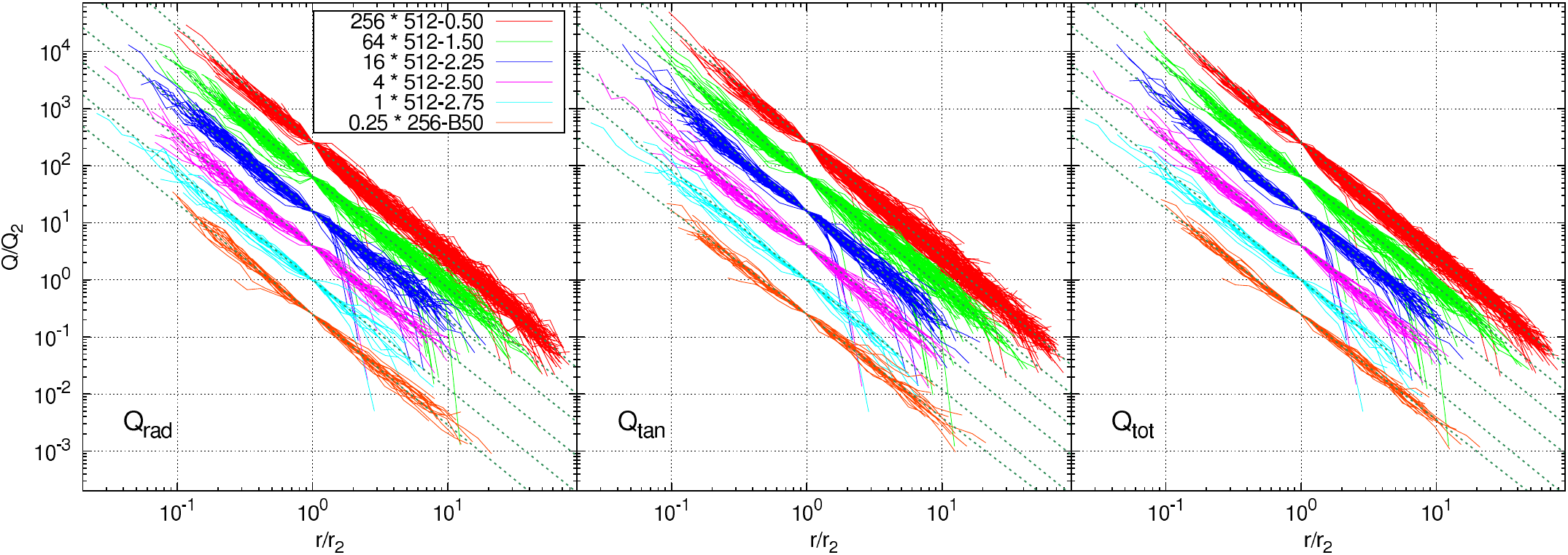}
  \caption{Phase-space density profiles $Q(r)$ for our sample of
  	haloes (radial, tangential, and total velocity dispersions
	are used in the left, middle, and right panel). The profiles of the
           different simulations are shifted on the $y$-axis for
           clarity, the factor by which they are shifted is given in the
           key.  For each simulation we also show the best-fitting
           $r^\alpha$ power-law, as given by the median value of
           $\alpha$ within the simulation; cf.
           table~\ref{tab:alpha_med_n} for the actual values.}
  \label{fig:comp_slopes}
\end{center}
\end{figure*}

In figure~\ref{fig:comp_slopes} the different $Q$-profiles
(i.e. radial, tangential and total) are shown in the normalization
described above alongside the best-fitting power-laws for the respective
simulation (i.e. the power law as given by the median of all $\alpha$'s
for a given model and $Q$, respectively).  The median slopes used to
draw the power laws are given in table~\ref{tab:alpha_med_n}, which we
will describe later in greater detail.

Furthermore, to avoid crowding, the profiles for the different
simulations are offset to one another by factors given in the key. For
all profiles the power law structure is quite convincing -- irrespective
of the simulation.  We also investigated the residuals of fits (though
not shown here), finding the average deviation per degree of freedom to
be on the order of 40 per cent, further strengthening the single power
law fit be a reasonable choice.  In the next subsection, we will use the
slopes derived for each halo.

\subsection{The universality of the slope}
\label{subsec:universality}

\begin{figure*}
\begin{center}
  \includegraphics[width=\textwidth]{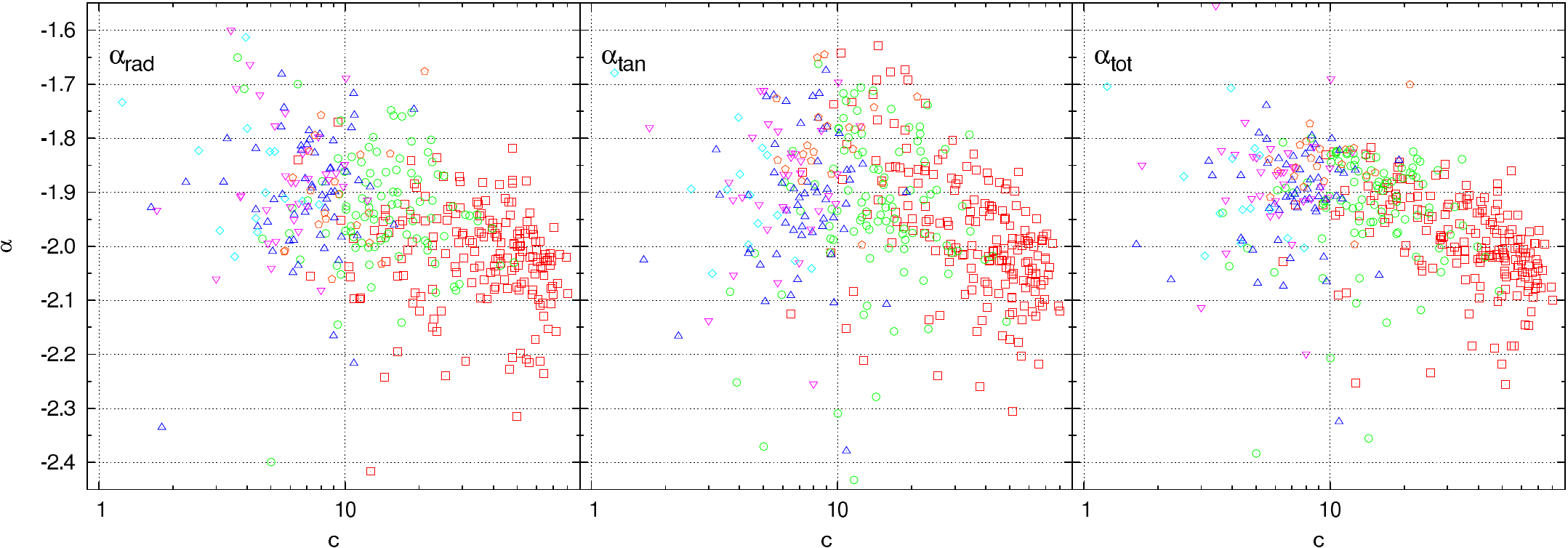}
  \caption{In this figure we show the derived slopes
           $\alpha$ of each halo versus its concentration
           $c=c_\mathrm{vir}/r_{2}$.  
           Each panel corresponds to using a different velocity
           dispersions to calculate $Q(r)$ (see equation~\ref{eq:qr}),
           the results for $Q_{\mathrm{rad}}(r)$ ($Q_{\mathrm{tan}}(r)$,
           $Q_{\mathrm{tot}}(r)$) are shown in the left (middle, right)
           column.  The haloes are colour coded
           according to the model they correspond to: red $n=-0.50$ 
           (open boxes), green $n=-1.50$ (open circles), blue $n=-2.25$
           (open upright triangles), magenta $n=-2.50$ (open downright
           triangles), cyan $n=-2.75$ (open diamonds) and for
           comparison haloes from a $\Lambda$CDM simulation are shown in
           orange (open pentagons).} 
  \label{fig:alpha_chi}
\end{center}
\end{figure*}

It is known that haloes formed in runs with different spectral index $n$
favour different concentrations\footnote{This is already
apparent in figure~\ref{fig:comp_slopes}: the $Q$-profiles are plotted
out to the virial radius and hence the final point is a direct measure
of concentration for our choice of units.}
but are otherwise compatible with a universal density slope
\citep[cf.][]{Knollmann2008}. We therefore investigate the slopes
$\alpha$ of individual haloes as a function of their concentration $c$.
This is shown in figure~\ref{fig:alpha_chi}, where the left (middle,
right) panel corresponds to the radial (angular, total) $Q$ profile.
Additionally, the simulations the haloes originate from are symbolised
by different colours (see the caption of figure~\ref{fig:alpha_chi} for a
description, we will use those symbols and colours for the remainder of
the paper).  The information contained in figure~\ref{fig:alpha_chi} is
hence two-fold:
a) the connection of $\alpha$ to the model\footnote{This means
effectively a dependence on the spectral index $n$ of the power spectrum
of initial density perturbations.} is visualised by the colour and b)
the dependence on the concentration is encoded on the abscissa.

\begin{figure*}
\begin{center}
  \includegraphics[width=\textwidth]{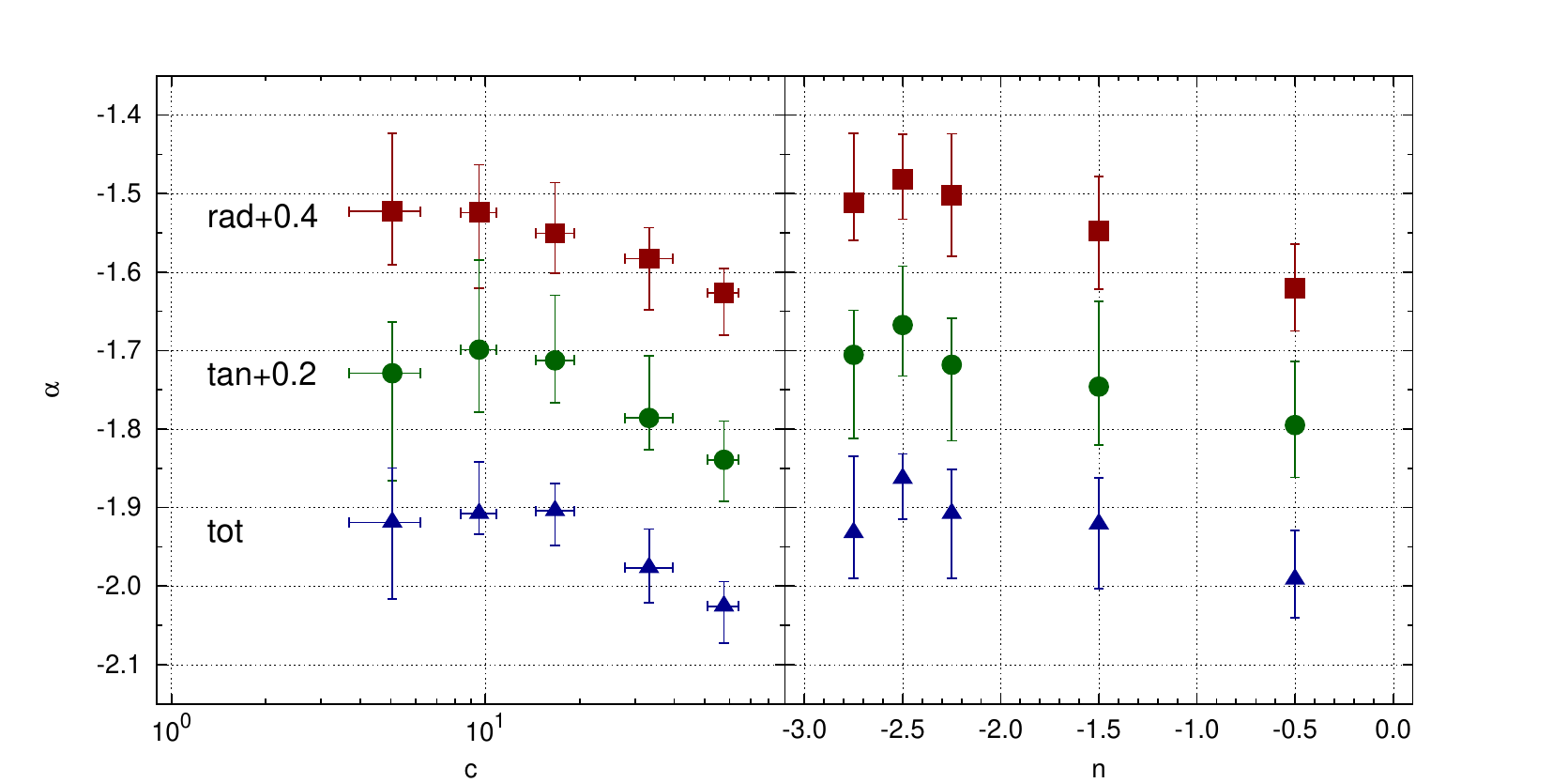}
  \caption{
In this figure we investigate the relation of the slope
$\alpha$ of the phase-space density profiles to the concentration $c$
(left panel) and the spectral index $n$ of the model (right panel).  
For clarity, the points corresponding to the tangential (dark green
filled circles) and radial (dark red filled squares)
$Q$-profile are offset to the ones corresponding to the total (dark blue
filled triangles) $Q$-profile by $0.2$ and $0.4$, respectively.  In the
left panel we have binned the $\alpha-c$ distribution shown in
figure~\ref{fig:alpha_chi} in $c$ regardless
of $n$, whereas in the right panel we combine the slopes of all haloes
within one model regardless of the concentration (see also
tables~\ref{tab:alpha_med_n} and \ref{tab:alpha_med_c}).  We show the
median and the lower and upper quartiles of each bin.}
  \label{fig:alpha_n_c}
\end{center}
\end{figure*}

To investigate those two effects more closely -- and to understand which
is the more fundamental dependence -- we will probe the relations separately,
looking for the one yielding the smaller scatter.  To this extent, we
bin all haloes from the scale-free models in $c$, regardless of their
model identity; the medians and the upper and lower quartiles in the
bins are shown in the left panel of figure~\ref{fig:alpha_n_c}.  We then
use the median $\alpha$ in each model to construct the right panel of
figure~\ref{fig:alpha_n_c}, showing the trend for $\alpha - n$.  The
curves for the different definitions of $Q$ are offset to one another to
avoid crowding.  The median values and the upper and lower quartiles are
also given in tables~\ref{tab:alpha_med_n} and~\ref{tab:alpha_med_c},
respectively.

We first note that, ignoring the model
identity and combing all haloes, a clear, yet noisy, trend can be
observed:  For haloes with a concentration $c \gtrsim 10$ the $Q$
profiles steepen with rising $c$, whereas a flattening of the $\alpha-c$
curve occurs for concentrations $c \lesssim 10$.  This is reflected in
the $\alpha-n$ relation, for flat $n$ models, we find steeper $Q$
profiles, with the $n=-2.75$ model, however, showing a steeper median
$\alpha$ than the $n=-2.50$ and $n=-2.25$ models.  We also note that
neither the $\alpha - c$ nor the $\alpha - n$ is clearly preferred as
the fundamental relation, both have a comparable scatter in $\alpha$,
with maybe the former slightly favoured.

In order to check whether our results are an artifact of the fitting
procedure we performed three additional test: a) we fitted the
$Q$-profiles confining the radial range to $0.1 < r/r_2 < 10$ (i.e. the
fitting has been performed over a fixed dynamical range for all the
haloes of all the different models considered here), b) we subdivided
the sample of each model into a low and high concentration bin and
calculated the median $\alpha$ in those two bins for each model, and c)
we calculated the reduced $\chi^2$ for each individual halo fit and
inspected its distribution. All these checks indicate that the results
seen and presented in figures~\ref{fig:alpha_chi}
and~\ref{fig:alpha_n_c} are stable.

For test a) we recover a
similar distribution as shown in figure~\ref{fig:alpha_chi}.  The
second test traces the global trend shown in the left panel of
figure~\ref{fig:alpha_n_c}, with the $n=-0.50$ and $n=-1.50$ models
showing a steeper $Q$-profile for the high concentration subsample,
whereas the $n=-2.25$ shows a similar $\alpha$ in both bins and the
steeper models show a steeper $\alpha$ for the low concentration bin,
e.g. the opposite of what is found in the flatter models. It should be
noted though, that due to the low numbers of haloes available in the
steep models ($n<-2.25$,  cf. table~\ref{tab:simus}), this might just be
a statistical effect.  Lastly, the probability distributions of
$\chi^2_{\mathrm{rad}}$, $\chi^2_{\mathrm{tan}}$, and
$\chi^2_{\mathrm{tot}}$ (not shown here though) confirm that all
profiles are fitted equally well with the assumption of a single
power-law, with the relative deviations per degree of freedom of the
order of 40 per cent.

\begin{table}
  \caption{Median slopes of the phase-space density profiles depending
           on the model.}
  \label{tab:alpha_med_n}
  \begin{tabular}{@{}ccccc}
    \hline
    Model & $\alpha_{\mathrm{rad}}$ & $\alpha_{\mathrm{tan}}$ &
    $\alpha_{\mathrm{tot}}$ \\
    \hline
    512-0.50 & $-2.02^{-1.97}_{-2.08}$ & 
               $-2.00^{-1.91}_{-2.06}$ &
               $-1.99^{-1.93}_{-2.04}$ \\
    512-1.50 & $-1.95^{-1.88}_{-2.02}$ &
               $-1.94^{-1.83}_{-2.02}$ &
               $-1.92^{-1.86}_{-2.00}$ \\
    512-2.25 & $-1.89^{-1.82}_{-1.98}$  &
               $-1.92^{-1.86}_{-2.02}$ &
               $-1.90^{-1.86}_{-1.99}$ \\
    512-2.50 & $-1.89^{-1.82}_{-1.93}$  &
               $-1.87^{-1.79}_{-1.93}$ &
               $-1.87^{-1.83}_{-1.92}$ \\
    512-2.75 & $-1.91^{-1.82}_{-1.94}$ &
               $-1.92^{-1.84}_{-2.02}$ &
               $-1.93^{-1.83}_{-1.99}$ \\
    256-B50  & $-1.94^{-1.83}_{-1.98}$ &
               $-1.81^{-1.75}_{-1.84}$ &
               $-1.85^{-1.82}_{-1.87}$ \\
    \hline
  \end{tabular}
\end{table}

\begin{table}
  \caption{Median slopes of the phase-space density profiles of the
           scale-free haloes depending on the concentration.}
  \label{tab:alpha_med_c}
  \begin{tabular}{@{}ccccc}
    \hline
    $c$ & $\alpha_{\mathrm{rad}}$ & $\alpha_{\mathrm{tan}}$ &
    $\alpha_{\mathrm{tot}}$ \\
    \hline
	$5.05^{6.20}_{3.69}$ &
	$-1.92^{-1.82}_{-1.99}$ &
	$-1.93^{-1.86}_{-2.07}$ &
	$-1.92^{-1.85}_{-2.02}$ \\
	$9.55^{10.83}_{8.34}$ &
	$-1.92^{-1.86}_{-2.02}$ &
	$-1.90^{-1.78}_{-1.98}$ &
	$-1.91^{-1.84}_{-1.93}$ \\
	$16.67^{19.24}_{14.46}$ &
	$-1.95^{-1.89}_{-2.00}$ &
	$-1.91^{-1.83}_{-1.97}$ &
	$-1.90^{-1.87}_{-1.95}$ \\
	$33.21^{39.61}_{27.81}$ &
	$-1.98^{-1.94}_{-2.05}$ &
	$-1.99^{-1.91}_{-2.03}$ &
	$-1.98^{-1.93}_{-2.02}$ \\
	$57.55^{64.01}_{51.00}$ &
	$-2.03^{-2.00}_{-2.08}$ &
	$-2.04^{-1.99}_{-2.09}$ &
	$-2.03^{-1.99}_{-2.07}$ \\
    \hline
  \end{tabular}
\end{table}

\subsection{The relation between the different $Q$ profiles}
\label{subsec:relation_q}

We will now investigate the relation of the different $Q$ profiles.  In
agreement with previous studies \citep[e.g.][]{Dehnen2005,
 Ascasibar2008} we find $Q_{\mathrm{rad}}$ to be steeper than
$Q_\mathrm{tot}$ in our $\Lambda$CDM sample, i.e.
($\alpha_{\mathrm{rad}}/\alpha_{\mathrm{tot}})_{\Lambda \mathrm{CDM}}
\approx 1.05$.  The scale-free simulations show at best a very
marginal trend of this kind: The slopes of the radial and total
$Q$-profiles for a given model tend to be much less different,
i.e. $(\alpha_r/\alpha_{\mathrm{tot}})_{\mathrm{scale-free}} \approx
1.01$.  However, the slope for a given $Q$-profile appears to change
across models, namely to drop with increasing spectral index $n$ (shown
in the right panel of \ref{fig:alpha_n_c}), with the notable exception
of the 512-2.75 model which favours steeper $Q$ profiles than the
512-2.50 model.  As we have shown above, this is also seen as a
$\alpha-c$ dependence, namely the slope of the $Q$ profiles first rises
with decreasing concentration $c$ until $c\approx 10$, from where on the
trend flattens and even turns around;  this is shown in the left panel
of figure~\ref{fig:alpha_n_c}.

\citet{Ascasibar2008} were the first to point out that not only
$Q_{\mathrm{rad}}$ and $Q_{\mathrm{tot}}$ are single power-laws, but
also $Q_{\mathrm{tan}}$.  They further showed that its slope is
shallower than the slope $Q_\mathrm{tot}$ basing their findings on a
suite of simulations of the concordance $\Lambda$CDM cosmology.  We
confirm this relation for the haloes in the (concordance) 256-B50
sample; we further find a single power-law to be a fair description of
the $Q_{\mathrm{tan}}$ profile in the scale-free samples.  However, in
the scale-free sample $\alpha_{\mathrm{tan}}$ tends to be slightly
steeper than $\alpha_\mathrm{tot}$, but still shallower than
$\alpha_{\mathrm{rad}}$.

\subsection{The velocity anisotropy of scale-free haloes}
\label{subsec:betaprofs}

\begin{figure*}
\begin{center}
  \includegraphics[width=\textwidth]{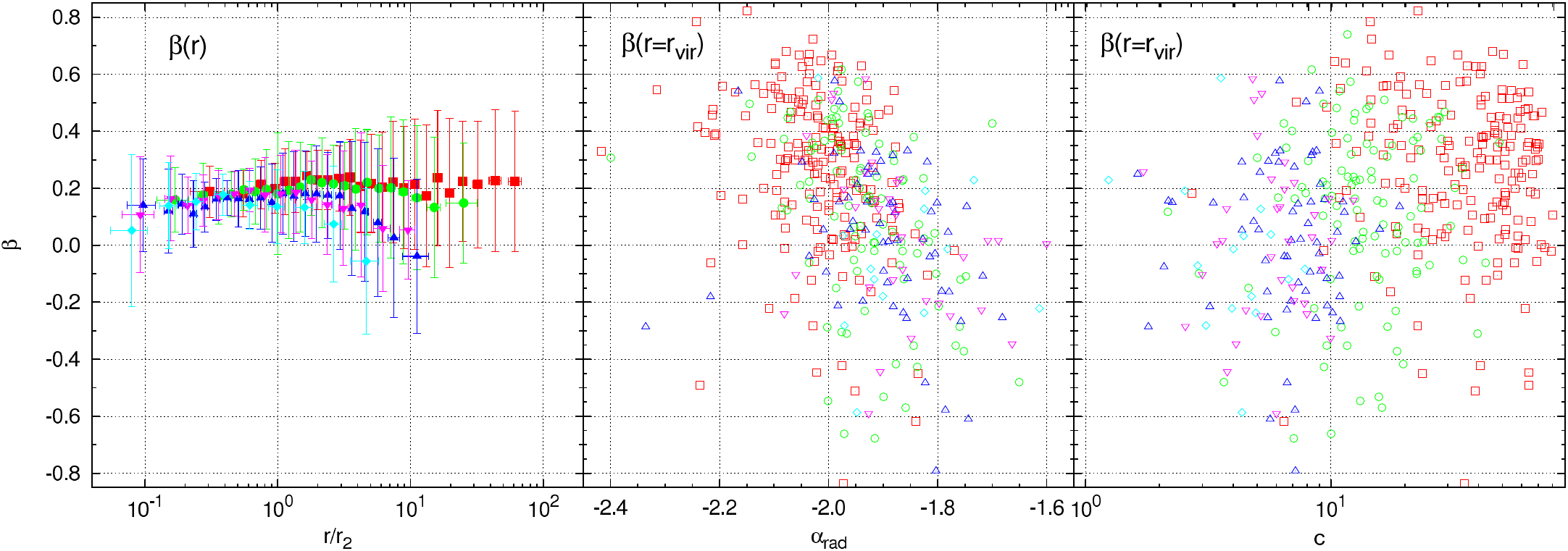}
  \caption{In the left panel, we show the mean $\beta$ profile in each
simulation, using the same colour coding as described in
figure~\ref{fig:alpha_chi}.  Note that the last point corresponds to
value of $\beta$ at the virial radius. The middle panel gives the value
of the  anisotropy profile at the virial radius,
$\beta(r_\mathrm{vir})$, as a function of the slope of the radial
$Q$-profile, $\alpha_\mathrm{rad}$.  In the right panel the distribution
of $\beta(r_\mathrm{vir})$ as a function of the concentration is
shown.}
  \label{fig:beta_all}
\end{center}
\end{figure*}

Given the universal density profile the difference between the power law
slopes of the different $Q(r)$ profiles implies that the anisotropy
$\beta$ of the velocity dispersion, given as
\begin{equation}
 \beta = 1- \frac{ \sigma^2_{\mathrm{tan}}}{2\sigma^2_{\mathrm{rad}}}
 \label{eq:beta_def}
\end{equation}
is not constant.  We investigate this in figure~\ref{fig:beta_all},
showing in the left panel the mean $\beta(r)$ profiles of the different
models. This shows an agreement of all profiles for $r\lesssim
r_\mathrm{2}$ but a dependence on the model in the outer part of the
haloes.  In the middle and right panel we now focus on the value of the
$\beta$ profile at the virial radius and check for how it correlates
with the slope of the radial $Q$ profile $\alpha_\mathrm{rad}$ and the
concentration $c$, respectively.

We see a trend for
$\beta(r_\mathrm{vir})$ to become larger for a steeper radial $Q$
profile, which also translates into a trend with $c$, namely, the larger
the concentration the larger becomes $\beta(r_\mathrm{vir})$; this is
shown in the right-most panel of figure~\ref{fig:beta_all}.  Recalling
equation~\ref{eq:beta_def}, we see that a larger $\beta(r_\mathrm{vir})$
means that the total velocity dispersion becomes more radial.  The
$\beta(r_\mathrm{vir}) - \alpha_\mathrm{rad}$ trend we observe agrees
with the findings of \cite{Zait2007} who calculated the $\beta$ profile
given the density profile and a power law $Q_{\mathrm{rad}}$ profile.
They showed that a more negative $\alpha_{\mathrm{rad}}$ implies a
larger $\beta$ for $r > r_2$, which means that the total velocity
dispersion is more radial.

\section{Summary \& Discussion}

We have analysed five high-resolution simulations of scale-free
cosmologies and investigated the logarithmic slope $\alpha$ of the
radial, angular and total phase-space density profiles
$Q(r)=\rho(r)/\sigma^3(r)$ when fitted by a single power-law $Q(r)
\propto r^\alpha$.  Additionally, we used a halo sample drawn from a
$\Lambda$CDM simulation as a control sample to verify our analysis
procedure and reproduced previous results.

Focussing on a well-defined sample of haloes in each scale-free
simulation, chosen to be in a comparable dynamical state, we
derived the slopes $\alpha_{\mathrm{rad}}$, $\alpha_{\mathrm{tan}}$
and $\alpha_\mathrm{tot}$ for each halo and computed the median slopes
in each simulation.  We found that the slopes for the three definitions
of the phase-space density profile $Q$ vary between the models, namely $Q$
becomes shallower with steeper spectral index $n$ of the initial power
spectrum $P(k) \propto k^n$, with the exception of the 512-2.75 model,
which shows flatter $Q$ profiles than found in the 512-2.50 model.  We
also find a trend for $\alpha$ to first become flatter with decreasing
concentration; this trend flattens at $c\approx 10$ and even turns
around for the tangential and total $Q$ profiles.  This is connected to
previous findings showing a correlation between the power spectrum index
$n$ and the concentration $c$.

But how can the $n$ - $c$ relation affect the $Q(r)$ profiles?  The scaled
density profiles appear to follow a universal form, independent of the
cosmological model~\citep[see e.g.][]{Navarro1996, Navarro1997,
Navarro2004}; in particular this has been shown explicitly for the
models studied here \citep{Knollmann2008}.  The dependence on the
cosmological model (i.e. the slope of the power spectrum of initial
density perturbations) is introduced via the concentration parameter, and
thereby by the dependence of the $\beta$ profile on the model.  The Jeans
equation, which dictates the structure of spherical DM haloes in
equilibrium, relates the density, phase space density and the velocity
anisotropy profiles \citep[e.g.][and references therein]{Zait2007}.
Zait et al. showed that for a given density profile a more radial
velocity dispersion at the outer parts of a halo implies a more negative
$\alpha_\mathrm{rad}$, which is precisely what we find and have shown in
figure~\ref{fig:beta_all}.  It is hence
interesting to note that, even though the density profile follows a
universal form, the $\beta$-profile does not, but rather depends on the
concentration \citep{Knollmann2008}.

Summarizing past and present results the  following conclusions can be
drawn about DM haloes in scale-free cosmologies. The density profile
shows a universal profile, upon an appropriate scaling, but the
concentration parameter depends on the power spectrum index $n$
\citep{Knollmann2008}.  Here it has been found the  phase-space density
profile obeys a power law.  The slope of its power law depends on whether
the phase-space is defined in terms of the radial, tangential or the
full velocity dispersion.  Furthermore, the slopes depends on the shape
of the power spectrum.  Also, the velocity anisotropy at
$R_\mathrm{vir}$ depends strongly on the power spectrum index.

The main problem addressed here is the universality of the phase-space
density profile, and a mixed answer has been found.  On the one hand it
has been very robustly shown that indeed a power law profile provides a
good fit in all models considered here.  Yet, the power law slope
$\alpha$ has been found to vary with power spectrum index $n$.  Given
our lack of understanding of the origin of the phase-space density power
law it is interesting to trace the reason for the dependence of the
slopes $\alpha$ on $n$.  It has been shown here that at least part of
that dependence is attributed to the concentration parameter variation
with $n$.  However, it is unclear whether there is a direct dependence
of the slope $\alpha$ on $n$.  This can be tested by comparing the slope
and $n$ relation for a sub-sample of haloes of different models with the
same value of the concentration parameter. Regrettably, the sample of
haloes used in the present work is too small to provide a clear answer.

\section*{Acknowledgements}

SRK and AK acknowledge funding through the Emmy Noether Programme by the
DFG (KN 755/1). Simulations and the analysis have been performed on the
sanssouci and luise cluster at the AIP, except the 256-B50 which was 
was performed on the Green machine at the Centre for Astrophysics and
Supercomputing at Swinburne University. We thank Chris Power for
providing us the snapshots.

\bibliographystyle{mn2e}
\bibliography{phase-space.bib}

\end{document}